\documentclass[RNAAS]{aastex631}
\usepackage{multirow}
\usepackage{amsmath}
\usepackage{footmisc}
\usepackage{hyperref}
\usepackage{comment}

\shorttitle{XDBS Catalog}
\shortauthors{Gobat et al.}

\begin{document}

\title{Catalog of X-ray Detected Be Stars (XDBS)}

\author{Caden Gobat}
\affiliation{Department of Physics, The George Washington University, 725 21st St NW, Washington, DC 20052}
\affiliation{Southwest Research Institute, 1050 Walnut Street, Suite 300, Boulder, CO 80302}

\author{Hui Yang}
\affiliation{Department of Physics, The George Washington University, 725 21st St NW, Washington, DC 20052}

\author{Oleg Kargaltsev}
\affiliation{Department of Physics, The George Washington University, 725 21st St NW, Washington, DC 20052}

\author{Jeremy Hare}
\affiliation{NASA Goddard Space Flight Center, Greenbelt, MD, 20771}
\affiliation{NASA Postdoctoral Program Fellow}

\author{Igor Volkov}
\affiliation{Department of Physics, The George Washington University, 725 21st St NW, Washington, DC 20052}


\begin{abstract}

We present a catalog of X-ray Detected Be Stars (XDBS) with 161 Be stars from the Be Star Spectra (BeSS) database having X-ray counterparts in the {\sl Chandra} Source Catalog v2.0, {\sl XMM-Newton} 4XMM-DR11 Catalog, or {\sl Swift} 2SXPS Catalog.
The  multi-wavelength catalog includes accurate optical positions, X-ray properties (fluxes, photon indices and hardness ratios), optical, near-infrared and infrared photometry, source classifications (when available), and other properties including proper motions, effective temperatures, X-ray to optical flux ratios as well.  We also provide a convenient \href{https://home.gwu.edu/~kargaltsev/XDBS/}{graphical user interface} which allows for easy visualization of the catalog content.

\end{abstract}

\section{Introduction} \label{sec:intro}
Be stars are hot, luminous B-type stars whose spectra show one or more emission lines. 
They have high rotational velocities (several hundreds of km\,s$^{-1}$, nearing the breakup limit) and possess equatorial decretion disks. 
Ultraviolet emission from the hot star ionizes the disk, which re-emits photons at longer wavelengths. 
Be stars are also interesting because they are often found in high-mass X-ray binaries  \citep[HMXBs;][]{2015A&ARv..23....2W} and $\gamma$-ray binaries \citep{2020mbhe.confE..45C}. The end products of Be star (and Be binary) evolution are not fully understood \citep{2019IAUS..346....1V} and may include exotic compact objects (e.g., magnetars).

While Be stars have been studied extensively in the optical band, studies of their X-ray properties have been more limited. 
We present the first catalog of X-ray Detected Be Stars (XDBS), available on GitHub at \url{https://github.com/huiyang-astro/XDBS/blob/main/master.csv}. 
An online interactive plotting tool \citep[described in][]{2021RNAAS...5..102Y} has also been built to visualize the multi-wavelength properties of the catalog, and can be accessed at \url{https://home.gwu.edu/~kargaltsev/XDBS/}.  

\section{Catalog description}

The catalog compiles multi-wavelength properties of 161 Be stars detected in X-rays as well as source classifications (e.g., HMXBs, $\gamma$ Cas analogs), when available. 
Sample of the catalog is shown in Table \ref{tab:XDBS}, as well as via GitHub. 
Below we describe the important details of the catalog construction.  

\subsection{Crossmatches}

We crossmatched  the Be Star Spectra catalog\footnote{There were 2264 entries in BeSS when we constructed this catalog. More Be stars are regularly added since BeSS is a living catalog.} \citep[BeSS;][]{2011AJ....142..149N} to {\sl Gaia} DR3 \citep{2022yCat.1355....0G} and {\sl Gaia} eDR3 distance catalogs \citep{2021AJ....161..147B} using a radius of 2\,$\arcsec$ to obtain accurate coordinates to be used in the subsequent crossmatching.
If the same BeSS star is matched to multiple {\sl Gaia} sources, only the closest match is kept. 
For BeSS sources that do not have {\sl Gaia} counterparts, the original BeSS coordinates are used for subsequent cross-matching. 

We then performed a cross-match with X-ray catalogs using the updated coordinates and error circles of 2\,$\arcsec$, 5\,$\arcsec$, and 9\,$\arcsec$ for the {\sl Chandra} Source Catalog v2 \citep[CSCv2;][]{2010ApJS..189...37E}, the {\sl XMM-Newton} 4XMM-DR11 Catalog \citep{2020A&A...641A.136W}, and the {\sl Swift}-XRT 2SXPS Catalog \citep{2020ApJS..247...54E}, respectively. The error circles correspond to the typical positional uncertainties of the respective X-ray catalogs. 
If multiple X-ray sources are found near the same BeSS star, only the nearest X-ray counterpart is kept. 
Eight sources were removed because they had either entirely missing/null flux values or only upper limit detections across all X-ray catalogs in which they were present.
We also removed 15 2SXPS detected sources with {\sl Gaia} Gmag$<9$ due to the large optical loading. 
We found 161 Be stars that are matched to at least one of the three X-ray catalogs with 74, 124, and 72 sources in CSCv2, 4XMM-DR11, and 2SXPS, respectively. 
We also applied a backward-matching of X-ray sources to {\sl Gaia} counterparts after we obtained the X-ray counterparts to verify the cross-matching. 
We also manually investigated all 161 matches and found that some BeSS optical coordinates needed to be updated so that the accurate {\sl Gaia} counterparts can be matched. We set a flag (\texttt{match\_flag}=1) to mark 20 questionable matches, which include the cases where the initial {\sl Gaia} counterparts of the Be stars do not match the {\sl Gaia} sources that the X-ray sources are backward-matched to, and/or those where we could not independently verify the BeSS coordinates. 
 There are two X-ray sources that do not have {\sl Gaia} counterparts which we flag with \texttt{match\_flag}=2.

Finally, to obtain near-infrared and infrared properties, we cross-matched our best-determined optical coordinates to the 2MASS \citep{2003yCat.2246....0C}, AllWISE \citep{2014yCat.2328....0C}, CatWISE2020 \citep{2021ApJS..253....8M}, and unWISE \citep{2019ApJS..240...30S} using a 1\,$\arcsec$ search radius.

\subsection{X-ray Fluxes} 
\label{sec:fluxconvert}

In order to systematically compare the X-ray fluxes and hardness ratios and to compensate for the differing energy band definitions of CSCv2, 4XMM-DR11, and 2SXPS, we converted them to common energy bands, using the CSCv2 definitions of soft (0.5--1.2\,keV), medium (1.2--2\,keV), and hard (2--7\,keV) bands as a standard. The 4XMM-DR11 and 2SXPS fluxes are converted using scaling factors calculated by assuming a power-law spectrum with photon index $\Gamma$.
We used the best-fitted values of $\Gamma$ from CSCv2 or 2SXPS where available (without including their uncertainties for the X-ray flux conversion), and otherwise took $\Gamma=1.7$  \citep[the assumption of the XMM catalog;][]{2009A&A...493..339W}.

We accounted for asymmetric uncertainties in our determinations of converted fluxes using a custom Python package (\texttt{asymmetric\_uncertainty}\footnote{\url{https://github.com/cgobat/asymmetric_uncertainty}}). 
We also replaced X-ray fluxes with values of zero in a given band with a very small value ($10^{-20}$\,erg\,s$^{-1}$\,cm$^{-2}$) to circumvent divide-by-zero errors during the X-ray flux conversion calculation. 

\subsection{Source Classifications} \label{sec:classification}

The known classifications of the X-ray sources in the XDBS catalog  include HMXBs based on \citet{2006A&A...455.1165L,2021A&A...647A.165D,2022arXiv220702114F} and SIMBAD \citep{2000A&AS..143....9W}, $\gamma$ Cas analogs \citep{2016AdSpR..58..782S,2018A&A...619A.148N,2020MNRAS.498.3171N}, and young stellar objects (YSOs) classified from SIMBAD if \texttt{main\_type} is \texttt{YSO}, \texttt{Orion\_V*}, or \texttt{Ae*}. All other sources are labeled as stars by default, but their true class could still be one of the above. 
The breakdown of 161 X-ray sources are: 48 HMXBs\footnote{These include 3 high-mass $\gamma$-ray binaries.}, 19 $\gamma$ Cas analogs, 12 YSOs and 82 stars. 

\section{Potential Applications}

The XDBS catalog is a useful tool for many potential applications, which include (but are not limited to) population studies of various types of X-ray sources, classifying unknown X-ray sources and building training datasets for machine-learning classification \citep[e.g.,][]{2022arXiv220613656Y}, and searching for rare type X-ray sources (e.g., $\gamma$-ray binaries).

\section{Acknowledgement}

This research is partly supported by Chandra X-ray Observatory award AR9-20005A and NASA ADAP award 80NSSC19K0576. 
This work has made use of the BeSS database (\url{http://basebe.obspm.fr}), and the VizieR catalogue access tool \citep{2000A&AS..143...23O}.

\begin{longrotatetable}
\begin{splitdeluxetable*}{cccccccccccccBcccccccccccccBcccccccccccccc}
\tabletypesize{\scriptsize}
\tablewidth{0pt} 
\tablenum{1}
\tablecaption{A subset of XDBS catalog.\label{tab:XDBS}}
\tablehead{
\colhead{Be\_star} & \colhead{Type} & \colhead{RA} & \colhead{DEC} & 
\colhead{Xcat} & \colhead{Xidentifier} &
\colhead{Gamma} & \colhead{Fb} & \colhead{Fs} & \colhead{Fm} & \colhead{Fh} & \colhead{HRms} & \colhead{HRhm} & 
\colhead{DR3Name} & \colhead{Plx} & \colhead{RPlx} & \colhead{PM} & \colhead{epsi} & \colhead{sepsi} & \colhead{RUWE} & \colhead{G} & \colhead{BP} & \colhead{RP} & \colhead{Teff} & \colhead{Gflux} & \colhead{dist} & 
\colhead{J} & \colhead{H} &  \colhead{K} & \colhead{W1} & \colhead{W2} & \colhead{W3} & \colhead{W4} &  
\colhead{Vsini} & \colhead{Vtran} & \colhead{LX} & \colhead{fX2O} & \colhead{match\_flag} & 
\colhead{Class} & \colhead{ref}  \\
\colhead{} & \colhead{} & \colhead{deg} & \colhead{deg} & 
\colhead{} & \colhead{} & \colhead{} &  \multicolumn{4}{c}{($\rm10^{-15}~erg~s^{-1}~cm^{-2}$)} & 
\colhead{}  & \colhead{} &
\colhead{} & \colhead{mas} & \colhead{} & \colhead{mas\,yr$^{-1}$} & \colhead{mas} & \colhead{} & \colhead{} & \colhead{mag} & \colhead{mag} & \colhead{mag}  & \colhead{K}  & \colhead{($10^{-12}$\,erg\,s$^{-1}$\,cm$^{-2}$)} & \colhead{kpc} & 
\colhead{mag}  & \colhead{mag}  & \colhead{mag}  & \colhead{mag}  & \colhead{mag} & \colhead{mag}  & \colhead{mag}  & 
\colhead{km\,s$^{-1}$} & \colhead{km\,s$^{-1}$} & \colhead{($10^{31}$\,erg\,s$^{-1}$)} & \colhead{($10^{-5}$)} & \colhead{} & 
\colhead{} & \colhead{} \\
\cline{8-11} 
} 
%\colnumbers
\startdata 
GSC 02342--00359 & B5e & 52.29341706 & 31.36639996 & CXO & J032910.3+312159 & $3.2\pm0.1$ & $284\pm11$ & $13\pm1$ & $77\pm3$ & $194\pm10$ & $0.72\pm0.05$ & $0.43\pm0.04$ & Gaia DR3 121406360248113920 & 3.3957 & 6.21 & $20.1\pm0.5$ & 2.901 & 1861.0 & 2.889 & $14.608\pm0.004$ & $15.45\pm0.01$ & $12.491\pm0.008$ &  \nodata  & $14.53\pm0.05$ & $0.31^{+0.06}_{-0.04}$ & $9.37\pm0.03$ & $7.99\pm0.03$ & $7.17\pm0.02$ & $6.37\pm0.01$ & $5.761\pm0.008$ & \nodata & \nodata &  \nodata  & $30^{+6}_{-4}$ & $0.33^{+0.13}_{-0.08}$ & $1955^{+73}_{-77}$ & 1 & YSO & 1\\
BQ Cam & Be & 53.74962972 & 53.17313998 & CXO & J033459.9+531023 & \nodata & $23\pm7$ & $1\pm1$ & $10\pm3$ & $12\pm7$ & $0.8\pm0.4$ & $0.1\pm0.3$ & Gaia DR3 444752973131169664 & 0.1343 & 6.67 & $0.52\pm0.02$ & 0.0 & 0.0 & 1.07 & $14.2\pm0.003$ & $15.475\pm0.005$ & $13.087\pm0.006$ &  \nodata  & $21.17\pm0.07$ & $5.6^{+0.7}_{-0.5}$ & $11.82\pm0.02$ & $11.21\pm0.03$ & $10.74\pm0.03$ & $9.97\pm0.02$ & $9.71\pm0.02$ & $9.08\pm0.03$ & $8.5\pm0.3$ &  \nodata  & $14^{+2}_{-1}$ & $9^{+4}_{-3}$ & $108\pm35$ & 0 & HMXB & 3\\
MEROPE & B6IVe & 56.58167145 & 23.94814381 & CXO & J034619.6+235653 & $8.7^{+0.9}_{-0.8}$ & $202^{+11}_{-10}$ & $180^{+9}_{-10}$ & $22\pm3$ & $0.00001^{+4.36435}_{-0.00001}$ & $-0.78\pm0.06$ & $-1.0\pm0.3$ & Gaia DR3 65205373152172032 & 7.067 & 24.69 & $50.1\pm0.2$ & 1.505 & 3081.0 & 2.33 & $4.173\pm0.004$ & $4.148\pm0.003$ & $4.159\pm0.005$ &  \nodata  & $217078\pm730$ & $0.143^{+0.006}_{-0.005}$ & $4.2\pm0.2$ & $4.3\pm0.2$ & $4.22\pm0.02$ & $4.3\pm0.2$ & $4.0\pm0.1$ & $4.16\pm0.01$ & $2.99\pm0.04$ & 240 & $34\pm1$ & $0.049\pm0.005$ & $0.093\pm0.005$ & 0 & star &  \nodata \\
Menkhib & O7.5IIIe & 59.74126439 & 35.79104207 & XMM & J035857.8+354728 & \nodata & $2494\pm5$ & $2053\pm5$ & $394\pm2$ & $47\pm1$ & $-0.678\pm0.002$ & $-0.786\pm0.006$ & Gaia DR3 219375904303684224 & 2.4457 & 9.35 & $2.7\pm0.2$ & 1.328 & 3186.0 & 2.236 & $3.975\pm0.003$ & $3.97\pm0.003$ & $3.889\pm0.004$ & 21760 & $260471\pm810$ & $0.42^{+0.06}_{-0.04}$ & $4.0\pm0.3$ & $4.1\pm0.2$ & $3.95\pm0.04$ & $4.0\pm0.3$ & $3.7\pm0.2$ & $3.95\pm0.01$ & $3.76\pm0.03$ & 213 & $5.4^{+0.9}_{-0.7}$ & $5.2^{+1.4}_{-1.0}$ & $0.957\pm0.004$ & 0 & star &  \nodata \\
lam Eri & B2IVne & 77.28660799 & -8.75409389 & XMM & J050908.8--084514 & \nodata & $78\pm6$ & $56\pm4$ & $15\pm2$ & $7\pm4$ & $-0.58\pm0.07$ & $-0.4\pm0.2$ & Gaia DR3 3182891931108590336 & 3.6256 & 15.63 & $2.9\pm0.2$ & 1.547 & 4507.0 & 2.283 & $4.243\pm0.003$ & $4.117\pm0.003$ & $4.386\pm0.005$ &  \nodata  & $203391\pm639$ & $0.27\pm0.02$ & $4.9\pm0.2$ & $4.83\pm0.08$ & $4.71\pm0.02$ & $4.7\pm0.2$ & $4.3\pm0.1$ & $4.07\pm0.01$ & $3.65\pm0.02$ & 318 & $3.7^{+0.4}_{-0.3}$ & $0.071^{+0.012}_{-0.01}$ & $0.038\pm0.003$ & 0 & star &  \nodata \\
\enddata
\tablecomments{See the entire table electronically. Column descriptions are provided at \url{https://github.com/huiyang-astro/XDBS/blob/main/XDBS_column_descriptions.pdf}.}
\end{splitdeluxetable*}
\end{longrotatetable}

\end{document}